\documentclass{emulateapj}

\usepackage{natbib}
\usepackage{epsfig}
\usepackage{apjfonts}

\newcommand{\rhill}{r_{\mathrm{H}}}

\newcommand{\figref}[1]{Figure \ref{#1}}
\newcommand{\eqref}[1]{equation (\ref{#1})}

\newcommand{\paperone}{Paper I\@}
\newcommand{\papertwo}{Paper II\@}
\newcommand{\TTW}{TTW\@}

\shortauthors{Jang-Condell \& Sasselov}

\begin{document}
\received{May 5, 2004}
\revised{September 28, 2004}
\accepted{October 8, 2004}
\slugcomment{To appear in the Astrophysical Journal}
\title{Type I Migration in a Non-Isothermal Protoplanetary Disk}
\shorttitle{TYPE I MIGRATION IN A NON-ISOTHERMAL PROTOPLANETARY DISK}
\author{Hannah Jang-Condell\altaffilmark{1}
and Dimitar D.~Sasselov}
\affil{Harvard-Smithsonian Center for Astrophysics\\
60 Garden St., Cambridge, MA 02138\\}
\altaffiltext{1}{Now at the Carnegie Institution of Washington, 
Department of Terrestrial Magnetism, 
5241 Broad Branch Rd., NW, Washington, DC 20015}
\email{hannah@dtm.ciw.edu}
\shortauthors{JANG-CONDELL \& SASSELOV}

\begin{abstract}
We calculate rates of Type I migration of protoplanets in a
non-isothermal three-dimensional protoplanetary disk, building upon
planet-disk models developed in previous work.  We find that including
the vertical thickness of the disk results in a decrease in the Type I
migration rate by a factor of $\sim2$ from a two-dimensional disk.
The vertical temperature variation has only a modest effect on migration rates
since the torques at the midplane are weighted heavily both because
the density and the perturbing potential are maximized at the
midplane.  However, temperature perturbations resulting from shadowing
and illumination at the disk's surface can decrease the migration rate
by up to another factor of 2 for planets at the gap-opening threshold
at distances where viscous heating is minimal.
This would help to resolve the timescale mismatch 
between the standard core-accretion scenario for planet formation 
and the survival of planets, and could help explain some of the 
rich diversity of planetary systems already observed.  
\end{abstract}


\section{Introduction}

The issue of planetary migration due to tidal interactions between
the planet and a circumstellar disk was first addressed by 
\citet{GT80}, who found that the potential of an orbiting planet 
excites Lindblad resonances in the disk, resulting in the transfer 
of angular momentum.  They found that the resulting torques on the 
planet are similar in magnitude, so that whether the planet 
moves inward or outward from angular momentum exchange 
depends on how the torques from the inner and outer disk balance out.  
\citet{ward86} found that given a decreasing radial 
temperature profile, the
torques from the outer disk are greater than those from the inner
disk, resulting in a net torque that forces the planet inward.  
Planetary migration via this mechanism is referred to as 
Type I migration to distinguish it from Type II migration, which is the 
radial drift of planets massive enough to clear all material out 
of the Lindblad resonances, thereby forming a gap in the disk 
and moving together with 
the disk as it is accreted onto the star \citep{wardA}.  

Subsequent work on two-dimensional disks 
confirms the prediction that planets in 
protoplanetary disks migrate inward on short timescales 
on the order of $\lesssim10^5$ yr for Earth mass planets 
\citep[e.g.,][]{artymowicz93b,korycanskypollack,wardA}.  
\citet[hereafter \TTW{}]{TTW} find that including the three-dimensional 
density structure of an isothermal disk 
increases the migration timescale by a 
factor of 2 or 3 over a two-dimensional calculation.  
\citet{menougoodman} calculate migration rates in 
a disk with a radial temperature profile expected in 
T Tauri disks and also find that the migration timescales
increase when the thickness of the disk is accounted for.  
In a disk model that includes MHD turbulence, the density fluctuations 
near the planet can be sufficient to change the magnitude and even 
the sign of the total net torque on a small planet, so that 
the planet migrates as a ``random walk'' rather than a steady 
inward drift \citep{nelsonpapaloizou4}.  

While the effect of the density structure in the vicinity of a 
protoplanet on its migration rate has been studied in detail both 
analytically and in numerical simulations, 
the effect of local temperature variations has not
\citep[e.g.,][]{kley99,kdh01,bate}.  
Most simulations of planets embedded in disks employ a vertically isothermal 
equation of state, so that the disk has a simple radially dependent 
temperature structure.
Similarly, the analytic calculations discussed above 
assume vertical isothermality for simplicity.  
In this paper, we analyze Type I migration torques from 
Lindblad resonances in a 
protoplanetary disk where viscous heating and stellar 
illumination are the primary heating sources.  Further, we examine 
how temperature variations induced by the presence of the planet itself 
affect Type I migration.  

Disk temperature variations can affect the net Lindblad torques by
changing the local pressure gradient in the disk.  
A pressure gradient shifts the angular frequency of the
gas orbiting the star with respect to Keplerian velocities, 
which in turn shifts the positions of the
Lindblad resonances \citep{wardA}.  
The forcing function depends on the distance to
the planet, so the torques on the planet vary with the locations of
the Lindblad resonances.  To date, no one has addressed how a
non-isothermal temperature structure will affect the magnitude of the
angular momentum transferred at the Lindblad resonances and how that
might affect migration rates.  A variation in temperature structure,
such as those described in \citet[hereafter \paperone{}]{paper1} and
\citet[hereafter \papertwo{}]{paper2}, due to radiative transfer
effects, may affect the balance of migration torques.

In this paper, we address how the temperature structure of the 
disk affects the locations of Lindblad resonances and the 
consequences for Type I migration.  
In \S\ref{tidal_torques} we derive the equations used for calculating 
the torques on the planet from Lindblad resonances in a non-isothermal disk. 
In \S\ref{disk_model} we describe our adopted model disk and 
briefly discuss how the presence of a planet changes the temperature 
structure of the disk.  
We present our results in \S\ref{results} and discuss their implications 
in \S\ref{discussion}.  
Finally, in \S\ref{summary} we present our conclusions.

\section{Torques at Lindblad Resonances}
\label{tidal_torques}

In a two-dimensional disk, only one wave mode is excited at each 
Lindblad resonance.  In a three-dimensional disk, vertical velocity 
perturbations are allowed, so an entire series of wave modes is 
excited at each Lindblad resonance 
\citep[\TTW{},][]{takeuchimiyama,98lubowogilvie}.
The lowest wave mode, referred to as the $n=0$ mode in \TTW{} and 
the $f$-mode in \citet{98lubowogilvie}, has
no vertical velocity component in a vertically isothermal disk, 
so it is often referred to as the two-dimensional wave mode.  
This corresponds to the wave that is 
excited in the two-dimensional disk, although its properties are 
somewhat different since it is propagated through a 
disk with a finite height instead of an infinitesimally thin disk.  
The bulk of the angular momentum, even in thermally stratified disks, 
is carried by the two-dimensional wave \citep{98lubowogilvie}, 
so we shall concentrate our analysis on this wave mode.  
Vertical motions do occur in a thermally stratified disk, but 
we assume that these are small compared to the horizontal velocity 
perturbations.  As it is the total angular momentum that the wave 
carries that interests us, the exact form of the wave propagation 
is outside the scope of this paper. 

The pressure gradient is important in the calculation of
relative torques because it determines the location of the resonances, 
indirectly contributing to the magnitude of the torque. 
Local temperature variations can shift the locations of the 
Lindblad resonances by changing the local pressure gradient.  
This so-called pressure buffer can significantly affect the 
total net torques from Lindblad resonances \citep{wardA}.  
We assume that 
the density $\rho$ and temperature $T$ are not necessarily constant 
with either radius or height, therefore the pressure $p$ and sound 
speed $c$ vary locally as well.

\subsection{Resonance Locations}

Euler's equation for the unperturbed disk is
\begin{equation}\label{euler}
\frac{\partial\mathbf{v}}{\partial t} + (\mathbf{v}\cdot\nabla)\mathbf{v}
= - \frac{1}{\rho}\nabla p 
	- \nabla\left(-\frac{G M_{\star}}{|\mathbf{r}|}\right),
\end{equation}
where we adopt an ideal gas equation of state so that 
\(p = \rho c^2 = \rho kT/\mu m_H\) where $k$ is the Boltzmann constant, 
$\mu$ is the mean molecular weight ($\mu=2$ for molecular hydrogen), 
and $m_H$ is an atomic mass unit.  
We adopt a cylindrical coordinate system, \(\mathbf{r} = (r,\theta,z)\) 
where $r$ is the distance from the $z$-axis and $\theta$ is the 
azimuthal angle.  Assuming that the gas in the disk travels 
in circular orbits with 
\( \mathbf{v} = r\Omega \hat{\mathbf{\theta}} \), 
then the radial component of \eqref{euler} becomes 
\begin{equation}
r\Omega^2 = \frac{1}{\rho}\frac{\partial p}{\partial r}
+ \frac{G M_{\star}}{r^2(1+z^2/r^2)^{3/2}}
\end{equation}
where $G$ is the gravitational constant, and $M_{\star}$ is the 
mass of the star.
The equation for the angular velocity of the disk is 
\begin{equation}\label{angvel1}
\Omega^2 = \Omega_K^2 \left(1+\frac{z^2}{r^2} \right)^{-3/2} +
\frac{c^2}{r^2}\left(\frac{\partial\log p}{\partial\log r}\right)
\end{equation}
where $\Omega_K = \sqrt{GM_{\star}/r}$.  
Defining 
\(k\equiv-\partial\log \rho/\partial\log r\),
\(l\equiv-\partial\log p/\partial\log T\), 
and $h\equiv c/\Omega_K$, we can rewrite \eqref{angvel1} as 
\begin{equation}\label{Omega}
\Omega^2 = \Omega_K^2 \left[1 - \frac{h^2}{r^2}\left(\frac{3z^2}{2h^2} 
+ k + l \right)\right],
\end{equation}
omitting terms of ${\cal O}(z^4/r^4)$ and higher.
This equation differs from calculations of the angular velocity 
in two-dimensional disks which use integrated quantities in that 
$\Omega$ varies with disk height \citep[e.g.,][]{wardA}.  
If $k_{\sigma} = -\partial\log\sigma/\partial\log r$ 
where $\sigma = \int_{-\infty}^{\infty}\rho\, dz$ is the surface density, 
then in general, $k\neq k_{\sigma}$.  
In the example of a vertically isothermal disk where 
\(\rho = \frac{\sigma}{\sqrt{2\pi}h} \exp(-z^2/2h^2)\),
the density gradient is 
\begin{equation}
k = k_{\sigma} +\left(\frac{z^2}{h^2}-1\right)
\left(\frac{l}{2}-\frac{3}{2}\right),
\end{equation}
which yields 
\begin{equation}
\Omega^2 = \Omega_K^2 \left\{1 - \frac{h^2}{r^2}\left[\frac{3}{2}
+ k_{\sigma} + \frac{l}{2}\left(\frac{z^2}{h^2}+1\right) \right]\right\}
\end{equation}
as in \TTW{}.  The angular velocity is constant with $z$ 
only in an isothermal disk, where $l=0$. 
Realistic protoplanetary disk models, however, are not isothermal.  

We now consider a planet orbiting the star in the plane of the disk 
at $r=a$ and with angular velocity $\Omega_p=\sqrt{GM_{\star}/a}$.  
The effective location of the $m$th order 
Lindblad resonance is where $r$ satisfies 
\begin{equation}
D_* \equiv \kappa^2 - m^2(\Omega-\Omega_p)^2 + (mc/r)^2 = 0
\end{equation}
where $\kappa$ is the local epicycle frequency of the disk, 
\begin{equation}\label{kappa}
\kappa = \frac{1}{r^3}\,\frac{\partial}{\partial r}(r^4\Omega^2)
\end{equation}  
\citep{artymowicz93,ward88,wardA}.  We assume that
$\kappa\approx\Omega$, since the departure from a Keplerian flow is
small, and we find that using \eqref{kappa} to solve exactly for
$\kappa$ yields very similar results for the resonance positions and
torques.  Defining \( \xi=mc/r\kappa \), then
\begin{equation}
\frac{\Omega_p}{\Omega} = 1+\frac{\epsilon}{m}\sqrt{1+\xi^2}
\end{equation}
where 
\begin{equation}
\epsilon = \left\{
\begin{array}{rl}
+1, & \mbox{for outer resonances}\\
-1, & \mbox{for inner resonances}.
\end{array}
\right.
\end{equation}
Then we can solve for the location of the Lindblad resonance, 
$\alpha_r=r/a$, as 
\begin{equation}\label{alpha}
\alpha_r^{3/2} = \left[1+\frac{\epsilon}{m}\sqrt{1+\xi^2}\right]
\left(\frac{\Omega}{\Omega_K}\right).
\end{equation}
We solve for $\alpha_r$ and $\xi$ 
iteratively using equations (\ref{Omega}) and (\ref{alpha}), as 
in \citet{wardA}.  
The iteration is necessary because the magnitude of the torque 
is sensitive to the precise position of the resonance.  
Simply calculating $\alpha_r$ to first order in $h/r$ can introduce 
a systematic shift in the resonance positions which will affect the 
overall balance of torques.  

\subsection{Disk Torque Strength}
\label{LRtorque}
In a two-dimensional disk, the magnitude of the torque exerted near
the $m$th order Lindblad resonance is
\begin{equation}
T_m = - \frac{\pi^2 m \sigma \Psi_m^2}{r\, dD_*/dr}
\end{equation}
where $\sigma$ is the unperturbed surface density of the disk. 
The forcing function of the planet, $\Psi_m$, is defined by \citet{wardA}
to be 
\begin{equation}
\Psi_m = \frac{ r\, d\phi_m/dr + 2mf\phi_m}{\sqrt{1+4\xi^2}},
\end{equation}
where $f = m(\Omega-\Omega_p)/\Omega$, 
and $\phi_m$ is the amplitude of the $m$th order Fourier 
component of the disturbing function, 
\begin{equation}
\phi_m = -\frac{Gm_p}{a} \, b^m_{1/2}(\alpha_r)
\end{equation}
where $b^m_{1/2}(\alpha_r)$ is known as the Laplace coefficient, 
defined as 
\begin{equation}
b^m_{1/2}(\alpha_r) = \frac{2}{\pi}\int_0^{\pi}
\frac{\cos m\theta\,d\theta}{\sqrt{1 - 2\alpha_r\cos\theta + \alpha_r^2}}.  
\end{equation}
At the effective location of the Lindblad resonance, 
$f=-\epsilon\sqrt{1+\xi^2}$. 

To truly calculate the torques at Lindblad resonances in a three-dimensional 
disk requires a solution to the linearized fluid equations in three 
dimensions.  For a polytropic or isothermal equation of state, the 
eigenmodes can be solved for directly; these cases have been studied 
by others \citep[][\TTW{}]{98lubowogilvie}.  The temperature 
structure of the disk calculated in our model is neither polytropic 
nor isothermal, so we require a numerical solution.  
To simplify the problem, we assume that the vertical velocity 
perturbations are small compared to the horizontal ones so that we can 
treat the disk as the sum of independent layers and add up the total 
torque from these layers, as suggested by \citet{ward88} and 
\citet{artymowicz93}.  This relies on the assumption that relatively 
little angular momentum is carried away by vertical velocity perturbations, 
so our calculations should be treated as a useful estimate of the 
three-dimensional Lindblad resonant torque rather than a
rigorous solution.

For a disk with vertical extent, the forcing function must be modified 
to account for the vertical averaging of the planet potential's disturbing 
function.  We can express the torque on the disk at height $z$ over 
an extent $dz$ as 
\begin{equation}\label{dT}
d\tilde{T}_m = - \frac{\pi^2 m \rho\tilde{\Psi}_m^2}{r\, dD_*/dr}\: dz 
\end{equation}
where $\rho$, $r$ and $D_*$ are evaluated at the location of the Lindblad 
resonance at height $z$ given the temperature structure at that height, 
and $\tilde{\Psi}_m$ is calculated with a modified Laplace coefficient 
\begin{equation}
b^m_{1/2}(\alpha_r,\zeta) = \frac{2}{\pi}\int_0^{\pi}\frac{\cos m\theta \, d\theta}{\sqrt{1-2\alpha_r\cos\theta + \alpha_r^2 + \zeta^2}}
\end{equation}
where $\zeta = z/r$.

As noted in \citet{GT80}, for $|\alpha_r-1|\ll 1$ and $m\gg 1$, 
most of the contribution to the integral comes from $\theta\ll 1$.  
In other words, local conditions around the planet have the 
most effect on angular momentum transfer at Lindblad resonances.  
This is because most of the angular momentum transfer occurs 
from close encounters between gas streamlines and the planet.  
We can then approximate the Laplace coefficient by replacing 
$\cos\theta$ with $1-\theta^2/2$ and extending the upper limit 
of the integral to infinity.  Then, 
\begin{equation}
b^m_{1/2}(\alpha_r,\zeta) \approx
\frac{2}{\pi\sqrt{\alpha_r}}\int_0^{\infty}
\frac{\cos \theta\,d\theta}
{\sqrt{m^2[(1-\alpha_r)^2+\zeta^2]/\alpha_r+\theta^2}}
=\frac{2}{\pi\sqrt{\alpha_r}}K_0(\Lambda)
\end{equation}
where $K_i$ is the modified Bessel function of order $i$ and 
\begin{equation}
\Lambda = m\sqrt{\frac{(\alpha_r-1)^2+\zeta^2}{\alpha_r}}.
\end{equation}
Making use of the relation $K_0'=-K_1$ \citep{abramowitzstegun}, 
\begin{eqnarray}
\frac{r}{GM_{\star}/a} \,\frac{d\phi_m}{dr} &=&
-\alpha_r\frac{d\,b^m_{1/2}}{d\alpha_r} \nonumber\\
&=& \frac{1}{\pi}\left\{\frac{K_0(\Lambda)}{\sqrt{\alpha_r}}
+ \frac{m [\alpha_r^2 - \zeta^2-1)] K_1(\Lambda)}
{\alpha_r\sqrt{(\alpha_r-1)^2+\zeta^2}}
\right\}.
\end{eqnarray}

We define a dimensionless forcing function 
\begin{eqnarray}
\psi_m &\equiv& \frac{\epsilon\pi}{2m}\,\frac{a}{GM_{\star}}\,
\sqrt{1+4\xi^2}\,\tilde{\Psi}_m \nonumber\\
&=& \frac{\pi}{2}\left[
-\epsilon\,\frac{\alpha_r}{m}\frac{d\,b^m_{1/2}(\alpha_r,\zeta)}{d\alpha_r}
+2\sqrt{1+\xi^2}\,b^m_{1/2}(\alpha_r,\zeta)
\right].
\end{eqnarray}
Substituting in the Bessel function formulation of $b^m_{1/2}(\alpha_r,\zeta)$,
\begin{equation}
\psi_m = 
\frac{\epsilon(\alpha_r^2 - \zeta^2-1)}
{2\alpha_r\sqrt{(\alpha_r-1)^2+\zeta^2}}\, K_1(\Lambda)
+\left(\frac{\epsilon}{2m}
+2m\sqrt{1+\xi^2}\right)\frac{K_0(\Lambda)}{\sqrt{\alpha_r}}.
\end{equation}
When $\zeta=0$, this equation reduces to the forcing function 
calculated by \citet{wardA} for a two-dimensional disk.  

The total torque integrated over the vertical extent of the disk 
is then 
\begin{equation}
\tilde{T}_m = 
\frac{\pi^2 m}{r\,dD_*/dr}\int_{-\infty}^{\infty}\rho\tilde{\Psi}_m^2\,dz .
\end{equation}
If we assume that the transferred angular momentum is entirely dissipated
in the disk, then the net torque exerted by the planet on the disk 
for the $m$th order inner and outer Lindblad resonances is 
\begin{equation}
\Gamma_m = \tilde{T}_m^{\mbox{\scriptsize outer}}
	-\tilde{T}_m^{\mbox{\scriptsize inner}}
\end{equation}
and the total torque is 
\begin{equation}
\Gamma_{\mbox{\scriptsize tot}} = \sum_{m>1} \Gamma_m.
\end{equation}
When $\Gamma$ is positive (negative), the planet loses (gains) 
angular momentum, and the planet migrates inward (outward).  

\subsection{Migration Rates}

Since we consider planets at or below the gap-opening threshold, we 
assume that the torques exerted on the disk by the planet are entirely 
dissipated in the disk and that the torque of the disk on the planet 
is equal to the torques from the planet on the disk.  We do not take 
into account the changes to the disk made by the action of the tidal 
torques.  We expect that in general, the effect of the torques will be 
to decrease the density of the nearby disk by pushing material 
out of the Lindblad resonances so our torque estimates are upper limits.  
In the absence of feedback form the disk, the 
rate of change in the planet's angular momentum is equal to the total 
torque, and the rate at which the planet moves is 
\begin{equation}
\frac{da}{dt} = -\frac{2 \Gamma_{\mbox{\scriptsize tot}}}{m_p a \Omega_p}. 
\end{equation}
Negative values of $da/dt$ indicate inward migration, positive 
values indicate outward migration.  The migration timescale is 
\begin{equation}
t_m = \frac{m_p\Omega_p}{\left|2\Gamma_{\mbox{\scriptsize tot}}\right|}
\end{equation}

\section{Disk Model}
\label{disk_model}

The model disk is described in detail in \papertwo{}, both with and 
without planets.  In this section we will summarize the features of the 
model disk and the temperature changes caused by the presence of planets.  
When calculating the positions and torque strengths of the Lindblad 
resonances, we do not take into account the response of the disk 
to torques from the planets.  In particular, we do not address 
density perturbations in the disk such as spiral density 
waves or gap-clearing for the larger planets in our models.  These 
density modifications will affect both the locations of the resonances 
as well as the magnitude of the torques, but that calculation is outside
the scope of this paper.

\subsection{The Unperturbed Disk}

To calculate the unperturbed disk structure, we adopt the formalism 
developed by \citet{calvet} and \citet{vertstruct,dalessio2}, 
with some simplifying assumptions.  
The model we employ is one of a gaseous disk rotating around a 
hot young star where dust is the primary opacity source, 
and the gas and dust are well mixed.  The star is no longer actively 
accreting and has no outflow, so the disk is quiescent.  
The star has temperature $T_{\star}=4000$ K, 
radius $R_{\star}=2\:R_{\sun}$, and
mass $M_{\star}=0.5\:M_{\sun}$.  
The disk is accreting at 
$\dot{M} = 10^{-8}\:M_{\sun}\,\mbox{yr}^{-1}$ and 
has a Shakura-Sunyaev viscosity parameter of 
$\alpha_{\mbox{\scriptsize SS}} = 0.01$ \citep{shaksun}.

To find the temperature structure of the model disk, 
we calculate radiative transfer through the disk with two heat
sources: viscous heating and stellar irradiation.  
\begin{figure*}[htbp]
\begin{center}
(a)\hspace{3.2in}(b)\hspace{3.2in}\mbox{}\\
\resizebox{3.2in}{!}{\includegraphics{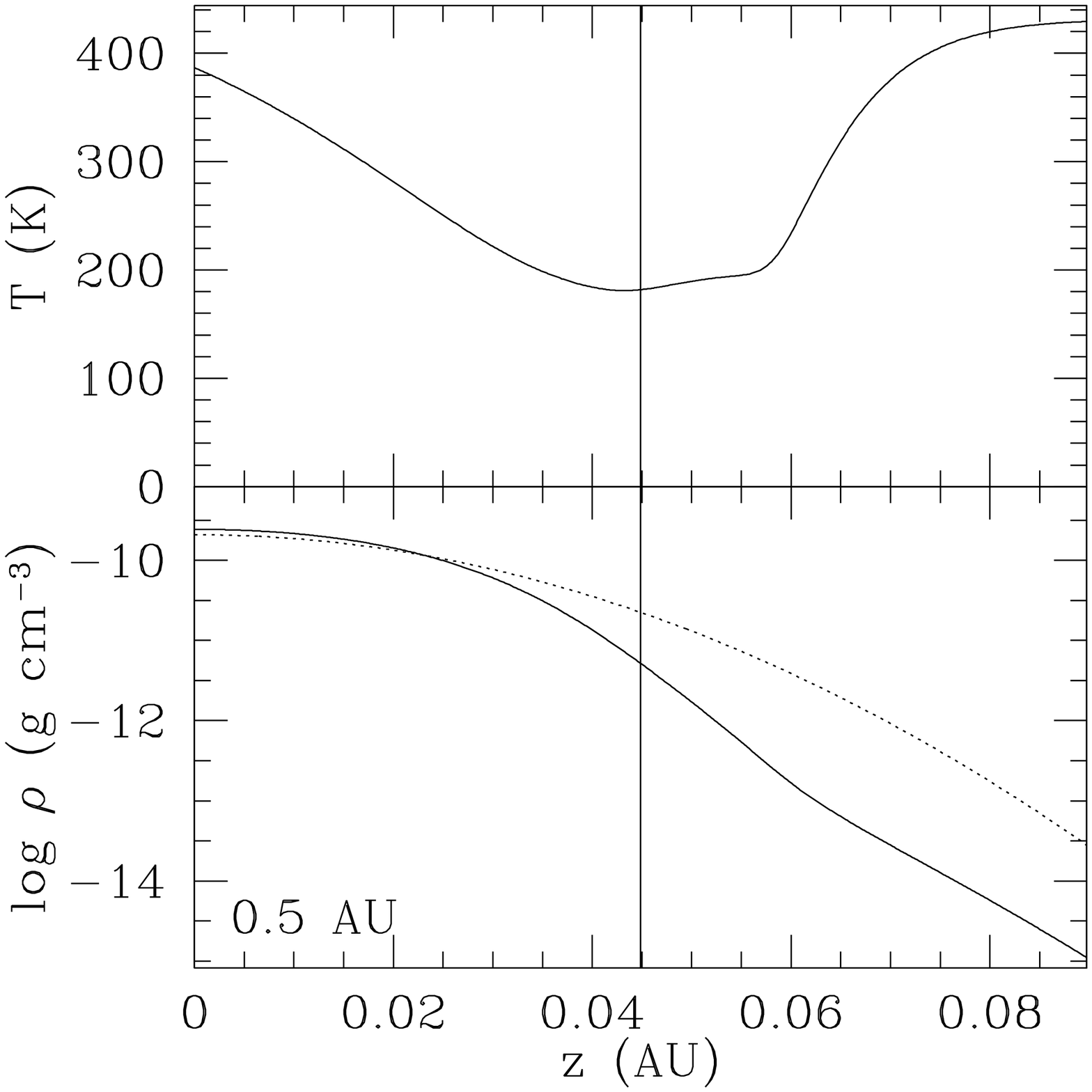}}
\resizebox{3.2in}{!}{\includegraphics{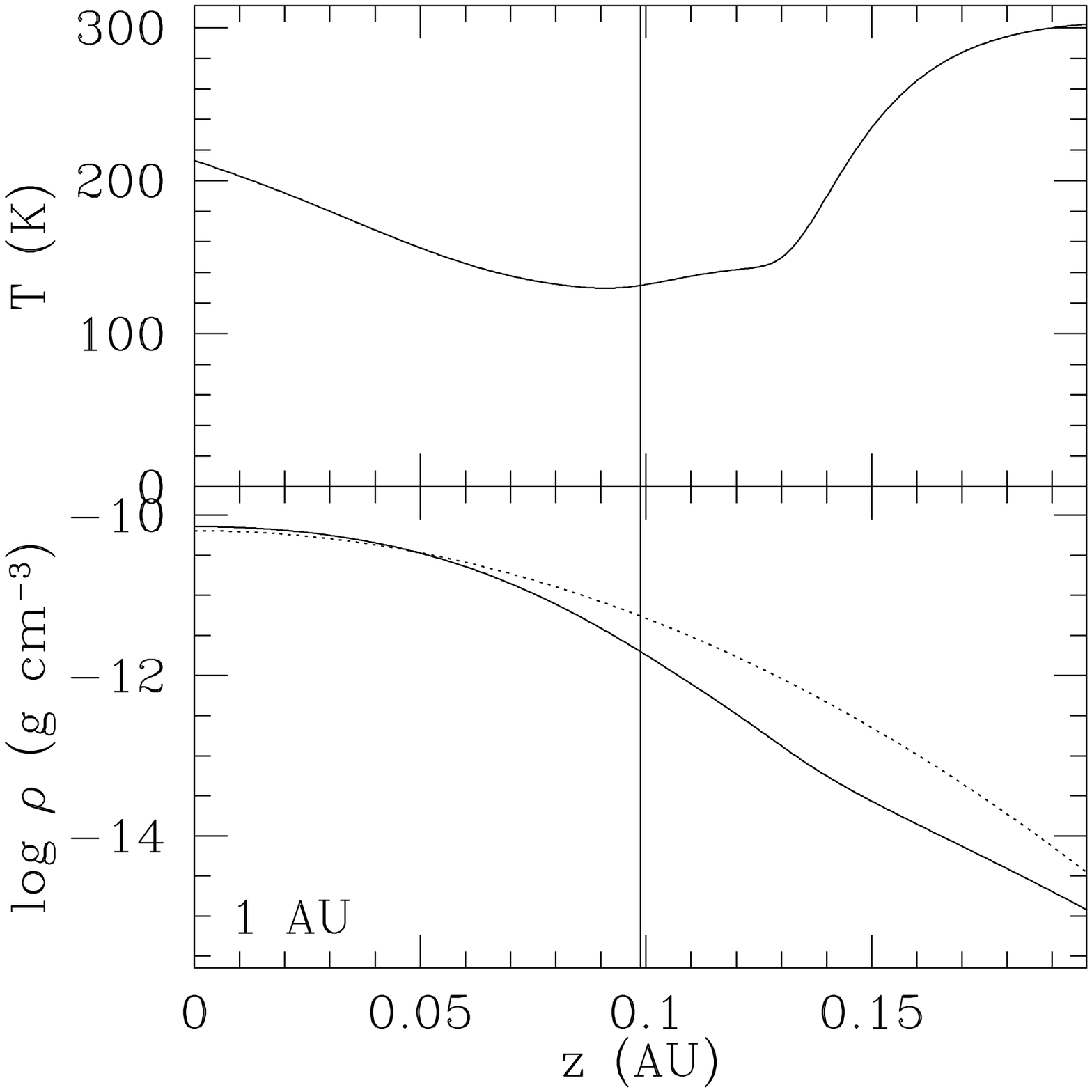}}\\
(c)\hspace{3.2in}(d)\hspace{3.2in}\mbox{}\\
\resizebox{3.2in}{!}{\includegraphics{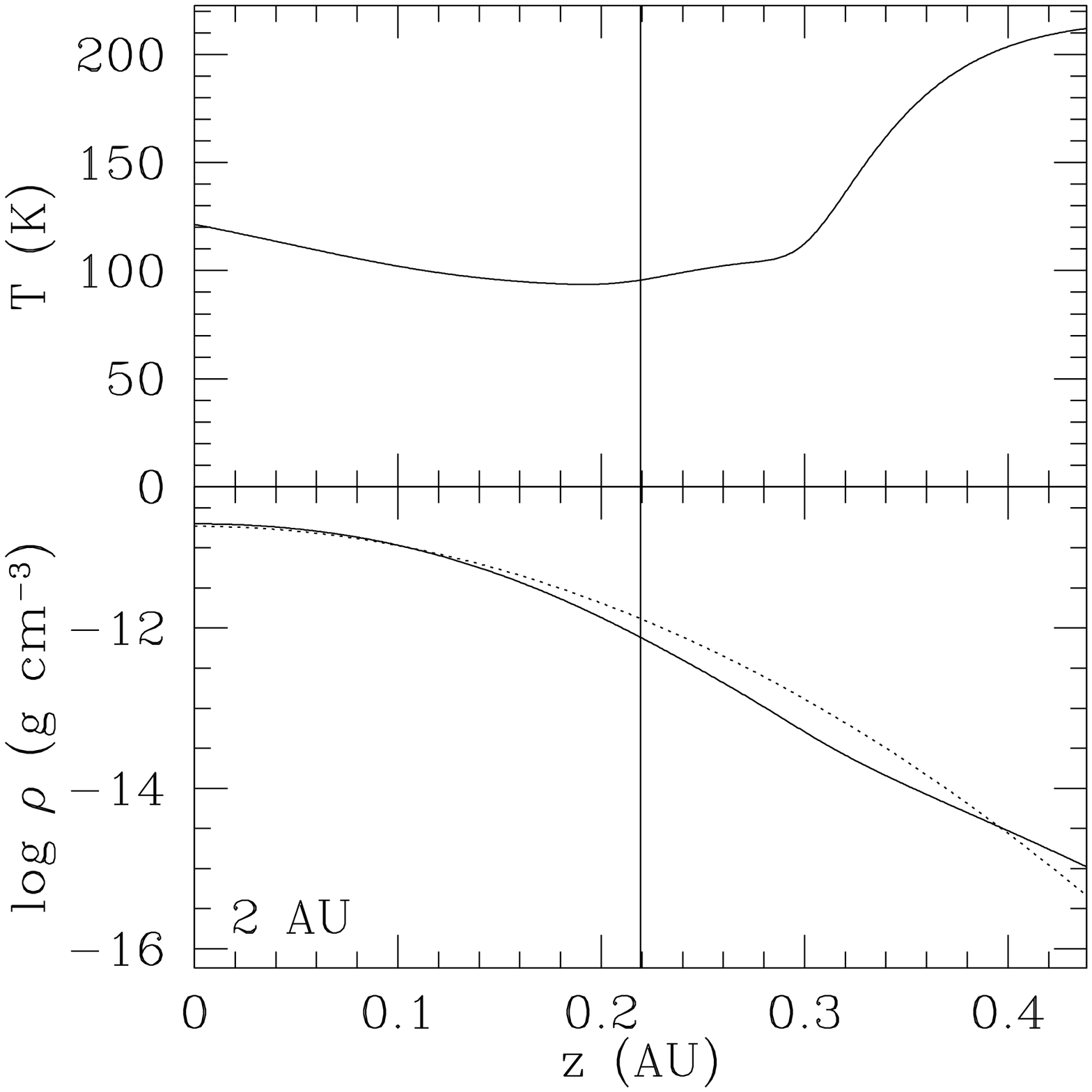}}
\resizebox{3.2in}{!}{\includegraphics{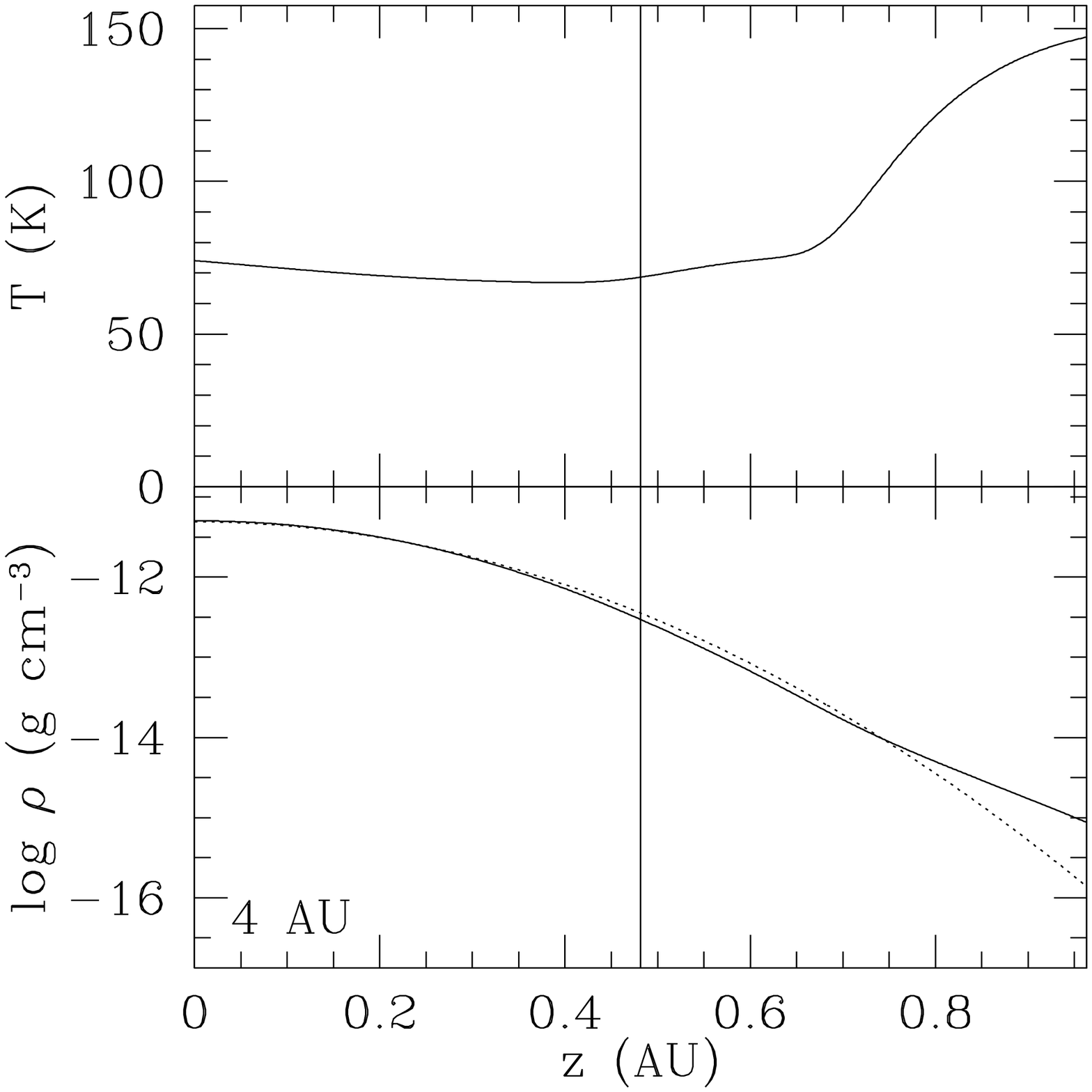}}
\end{center}
\caption{\label{diskprof}
Vertical temperature and density profiles of the unperturbed disk 
at (a) 0.5, (b) 1, (c) 2, and (d) 4 AU.  
The vertical line in each plot shows 
the location of the photosphere.  The dotted line shows the density 
profile of a vertically isothermal disk at the temperature of the 
midplane with the same surface density for comparison.
}
\end{figure*}
\figref{diskprof} summarizes the resulting vertical 
temperature and density profiles.  The upper and lower panels of each 
plot shows the temperature and density, respectively, 
of the disk versus scale height at 0.5, 1, 2, and 4 AU.  
The dotted line in the lower panels show the density profile 
for a vertically isothermal disk with the same midplane 
temperature and surface density for comparison.  The vertical 
line in each plot indicates the location of the photosphere, 
where the optical depth of the disk is $2/3$.  
Viscous heating is dissipated at the midplane, 
so the temperature decreases with increasing $z$.  
Stellar irradiation at the surface creates a temperature inversion, 
where the uppermost layers are hotter than lower layers.  
The photosphere is approximately the location of the vertical temperature 
minimum.

The viscous temperature goes as $r^{-3/4}$ while 
the stellar radiation temperature goes as $r^{-1/2}$, so that close 
to the star, viscous heating dominates.  
This causes the midplane temperature to be 
much higher than even the surface temperature at small $r$. 
Conversely, at larger radii 
the midplane temperature becomes close 
to the photosphere temperature.

\subsection{Disk with a Planet -- Spatial Temperature Variations}

When we insert a planet into the model disk, its 
gravitational potential induces a compression 
of the disk material near it, resulting in a decrement in the density 
at the disk's surface.  Thus, an isodensity contour at the height 
of the photosphere takes on the shape of a well.  When this well is 
illuminated by stellar irradiation at grazing incidence, 
it results in cooling in a shadowed 
region and heating in an exposed region.  

The temperature variations caused by shadowing and illumination 
effects can change the local pressure gradient, which will 
in turn shift the locations of the Lindblad resonances.  
As noted in \S\ref{LRtorque}, the integral for the 
perturbing potential is weighted toward $\theta\ll 1$ for 
$|\alpha_r-1|\ll 1$ and $m\gg 1$, implying that disk properties 
immediately around the planet have the 
most effect on angular momentum transfer at Lindblad resonances.  
Although the equations for calculating torques at Lindblad 
resonances assume an axisymmetric disk structure, we can set 
quantities such as density and temperature equal to the local 
values rather than azimuthally averaged quantities.  

In order to quantify the effect of temperature variations on 
Lindblad torques, we extract the temperature structure through a 
slice of the planet-disk models calculated in \papertwo{}, 
perpendicular to the planet's orbit and through the position 
of the planet and use this to determine 
the perturbed pressure gradient near the planet.  This yields 
new values for the tidal torques on the disk, and correspondingly 
different migration rates.  Our calculations are meant to put limits 
on the effect temperature variations from radiative transfer effects 
might have on migration rates of planets embedded in disks.  

We use a spline interpolation to calculate the temperature between the
gridded points, since we want both $T$ and $dT/dr$ to be continuous
and smooth.  Since the density structure within about $\rhill$ of the
planet is not well described because this material is being accreted
onto the planet, the temperature structure in this region is also not
well understood in our models.  For this reason, we assume that the
temperature simply varies linearly across the region within $\rhill$
from the planet.  This assumption does not significantly affect the
calculation of the tidal torques, since the effective resonance locations 
are at $|r-a|\gtrsim 2h/3$.

\section{Results}
\label{results}

In \papertwo{}, we calculated the three-dimensional temperature 
structure around planets embedded in protoplanetary disks, 
taking shadowing and illumination effects at the surface of the disk 
into account.  We shall put limits on the effect that these 
temperature variations can have on migration by taking a 
radial slice through the calculated volume at the planet's 
position and finding the 
migration rates associated with these temperature profiles.  
We calculate the torques at the Lindblad resonances and integrate 
in $z$ from the midplane to the photosphere.  Note that the photosphere 
is typically about twice the scale height of the disk above 
the midplane.  Above the photosphere, 
the disk is very rarefied, more than an order of magnitude less in
density, so contributions to the total torque are not significant.  

\begin{figure}[htbp]
\plotone{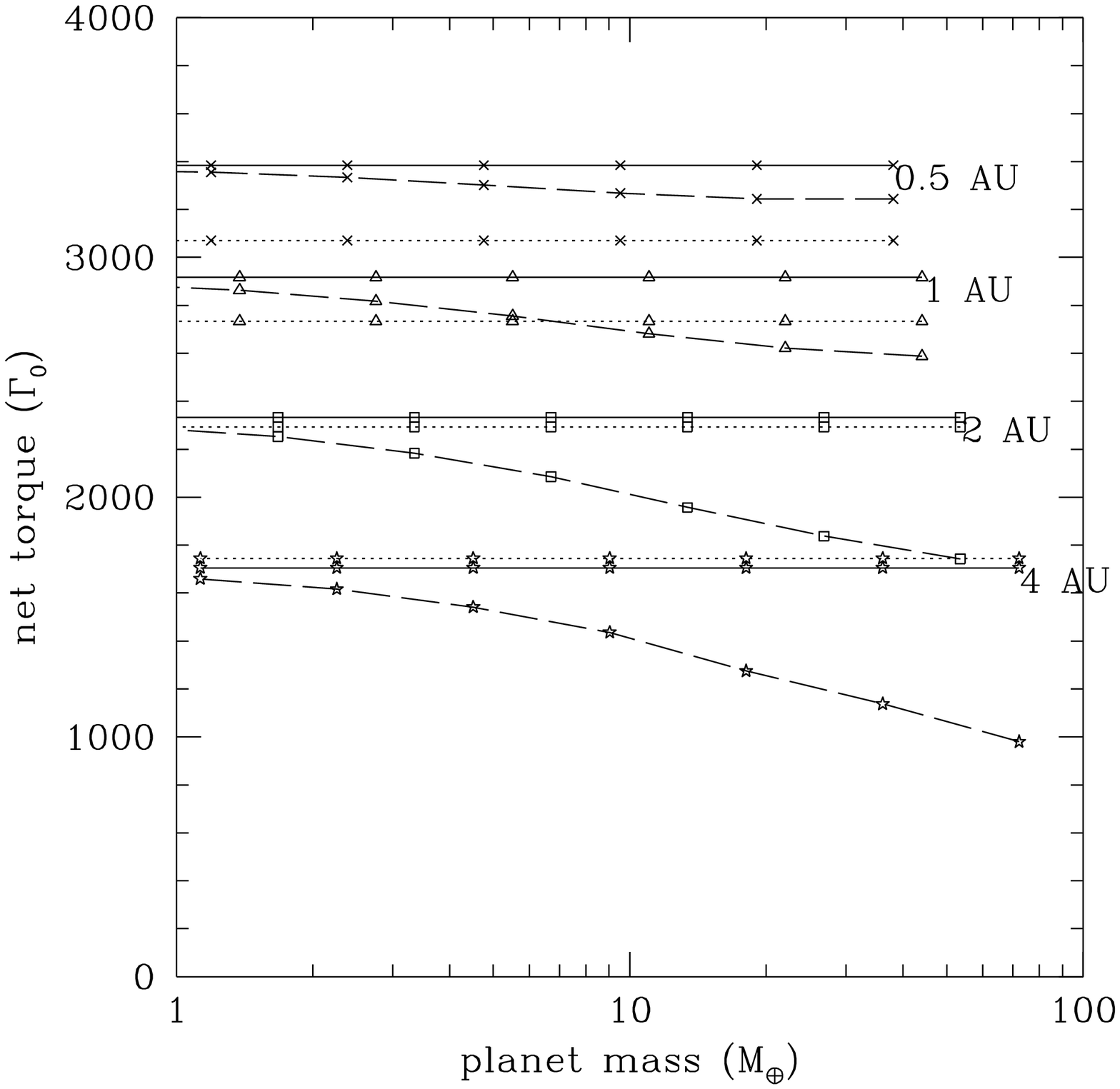}
\caption{\label{results_summary}Total net torques at 
0.5 (crosses), 1 (triangles), 2 (squares) and 4 (stars) AU.  
The solid lines indicate the total net torque for the unperturbed model disk 
and the dotted lines indicate the total net torque for an 
isothermal disk at the midplane temperature.
The dashed lines indicate the total net torque for the model 
disk with a planet of the mass indicated by the horizontal axis.  
}
\end{figure}
\figref{results_summary} summarizes the total net 
torques calculated for varying planet masses at several distances: 
crosses for 0.5 AU, triangles for 1 AU, squares for 2 AU, 
and stars for 4 AU.  The torque is normalized to 
\begin{equation}
\Gamma_0 = \left(\frac{m_p}{M_{\star}}\right)^2 \sigma_0 a^4 \Omega_p^2.
\end{equation}
Note that this quantity incorporates the scaling with both stellar 
and planetary masses.  
The solid lines show the torques for the unperturbed model disk, and the 
dotted lines indicate the torques for a vertically isothermal disk 
with the same integrated surface density with temperature 
equal to that at the midplane.  The dashed lines show how 
the total net torque varies with planet mass when we include the 
temperature variations from radiative transfer.  

In order to better understand how temperature perturbations 
alter migration torques, we discuss how the torque on the disk varies 
with $z$.  Noting that
\begin{equation}
\Gamma_{\mbox{\scriptsize tot}} = \sum_{m>1} \int_{-\infty}^{\infty} 
\left(\frac{d\tilde{T}_m^{\mbox{\scriptsize outer}}}{dz}
- \frac{d\tilde{T}_m^{\mbox{\scriptsize inner}}}{dz}\right)\,dz
\end{equation}
we define the scaled torque density as 
\begin{equation}
\gamma(z) = \frac{\sigma_0}{\rho}\frac{d\Gamma_{\mbox{\scriptsize tot}}}{dz}
= \frac{\sigma_0}{\rho}
\sum_{m>1} 
\frac{d\tilde{T}_m^{\mbox{\scriptsize outer}}}{dz}
- \frac{d\tilde{T}_m^{\mbox{\scriptsize inner}}}{dz}
\end{equation}
Since $\gamma(z)$ is scaled by $\sigma_0$, this quantity has the 
units of torque.  

\begin{figure}[htbp]
\plotone{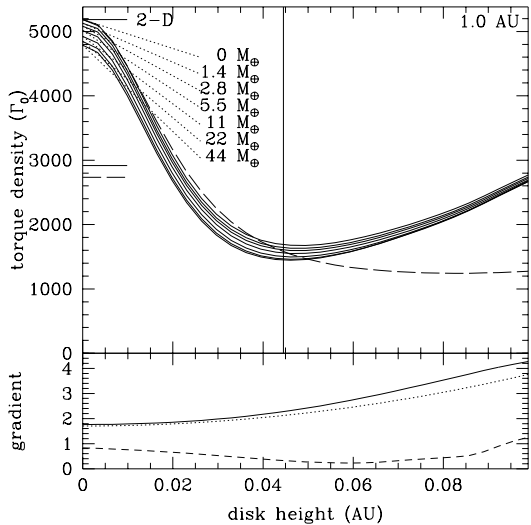}
\caption{\label{t1}Scaled torque density profiles, and 
density and temperature gradients as a function of height at 1 AU. 
The vertical line indicates the disk's scale height.
({\em Upper panel}): The scaled torque density, in terms of $\Gamma_0$, 
is plotted versus height from the midplane 
for a disk with and without a planet, the mass of the planet 
indicated by the labels.  The dashed line indicates the net torque 
in a vertically isothermal disk at the midplane temperature.  
The short horizontal lines on left edge of the plot show 
the total integrated torque for the unperturbed ({\em solid line}) 
and vertically isothermal ({\em dashed line}) disk models.  
({\em Lower panel}): The temperature gradient, $l$ is plotted 
as a dashed line.  
The quantities \([k+3(z\Omega_K/c)^2/2]\) ({\em solid line})
and \([k_{\sigma}+3/2+(l|_{z=0})z^2/2h_0^2]\) ({\em dotted line})
are also plotted.  
}
\end{figure}
\begin{figure}[hbtp]
\plotone{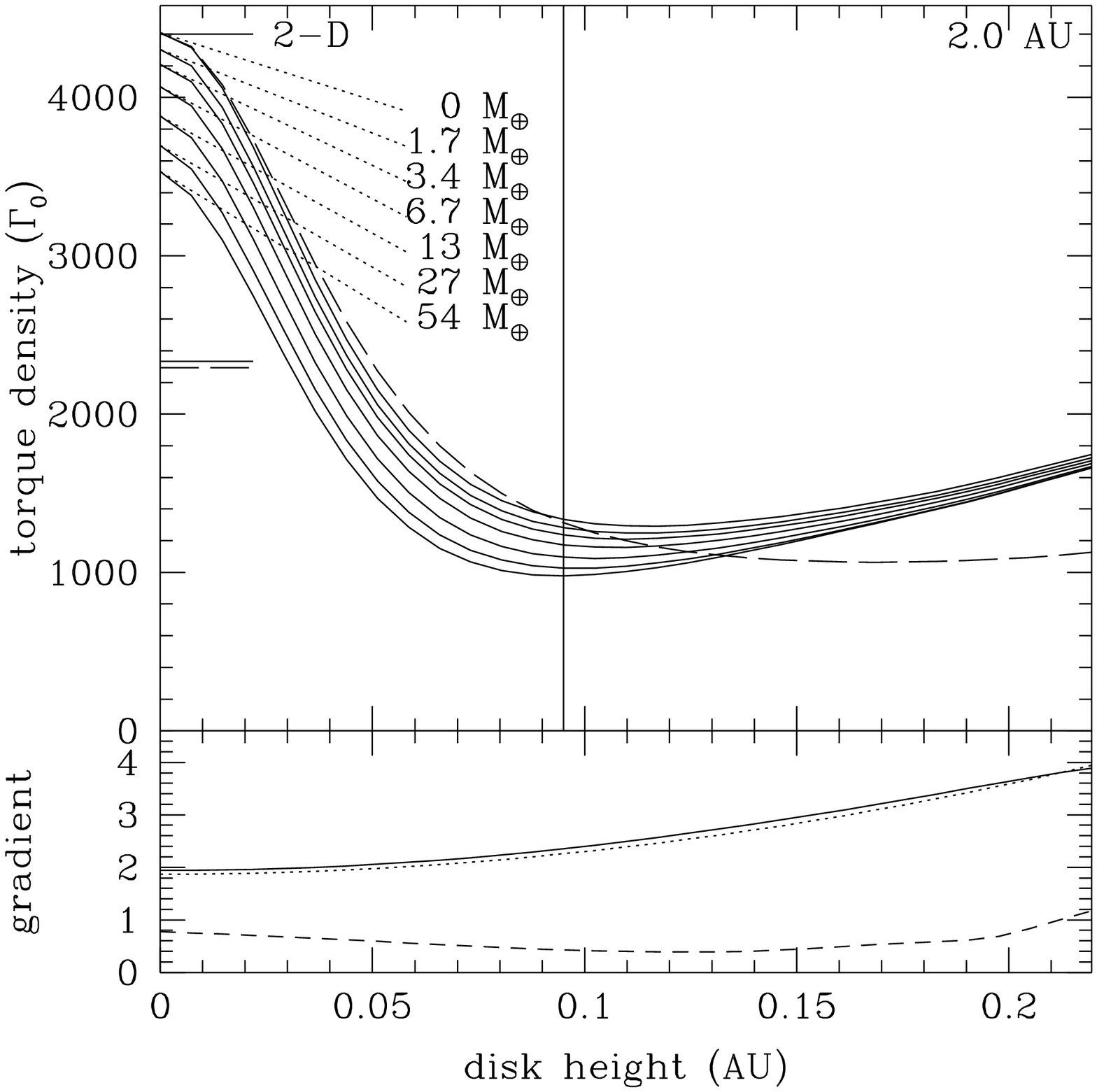}
\caption{\label{t2}Scaled torque density profiles, and 
density and temperature gradients as a function of height at 2 AU.  
See \figref{t1} for a full description of this plot.}
\end{figure}
\begin{figure}[hbtp]
\plotone{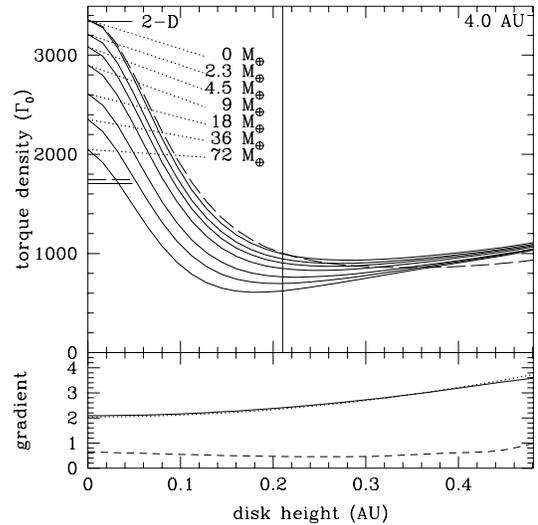}
\caption{\label{t4}Scaled torque density profiles, and 
density and temperature gradients as a function of height at 4 AU.  \
See \figref{t1} for a full description of this plot.}
\end{figure}
We plot the scaled torque density in terms of $\Gamma_0$
versus $z$ at 1, 2, and 4 AU in the upper panels of 
Figures \ref{t1}, \ref{t2} and \ref{t4}, respectively.  
In each plot, the vertical line indicates the scale height of the disk, 
$h_0=c_0/a\Omega_p$ where $c_0$ is the sound speed at the midplane 
of the disk.  
The short lines on the left edge of the plot show 
the total integrated torque in terms of $\Gamma_0$ for the 
unperturbed model disk ({\em solid line}) and a vertically 
isothermal model ({\em dashed line}).  The line labeled ``2D'' shows 
total net torque predicted for a two-dimensional disk with 
the same surface density profile as the model disk, and
temperature given by the midplane of the disk.  Note that this 
value is almost exactly equal to the value of the scaled torque 
density at the midplane for a disk with no planet.  
The solid lines indicate the torque density profiles for the
disk with temperature perturbations induced by planets of varying
mass, as labeled on the plot.  The topmost line, labelled 
0 M$_{\oplus}$, shows the scaled torque density for the unperturbed
model disk.  For comparison, we have plotted the scaled torque
density profile for a vertically isothermal disk with $T = T(z=0)$ 
as a dashed line.

Close to the midplane, the torque density falls off rapidly with $z$.  
It then levels off and rises again toward high altitudes.  
The sharp drop-off occurs because of the dependence on $z$ of the 
perturbing potential, 
indicating that the thickness of the disk does play an important 
role in diluting angular momentum transfer at Lindblad resonances.  
All the profiles follow the same basic shape near the midplane, 
including the vertically isothermal one.  

The torque density for the model disk 
increases at high altitudes above that of the isothermal disk 
because of several attributes of the disk that all increase 
the relative differential torque higher in the disk.  
Firstly, the Lindblad torque scales roughly as $c^{-2}$ 
\citep{ward86,korycanskypollack} and the vertical 
temperature minimum roughly corresponds to the 
photosphere, which is the limit of our integration in $z$.  
Secondly, the radial temperature gradient, $l$, rises slightly between 
the disk scale height and the photosphere.  We plot 
$l$ versus $z$ as a dashed line in the lower panel 
in each of Figures \ref{t1}, \ref{t2}, and \ref{t4}.  
The higher values of $l$ favor torques from the outer 
Lindblad resonances by shifting their locations, so the net 
torque rises \citep{wardA}.  Thirdly, the density gradient, $k$,  
is greater than the isothermal density gradient.  
As shown in \eqref{Omega}, the location of the resonance is shifted 
by a factor proportional to $(3z^2/2h^2+k+l)$.
The quantity \([k+3(z\Omega_K/c)^2/2]\) in the model disk 
is plotted as a solid line 
in the lower panel of Figures \ref{t1}, \ref{t2}, and \ref{t4}.  
For comparison, 
\([k_{\sigma}+3/2+(l|_{z=0})z^2/2h_0^2]\), 
which is the equivalent quantity in a vertically isothermal disk, 
is also plotted as a dotted line.  The solid line rises above the 
dotted line most noticeably at 1 AU in \figref{t1}, which is 
also where the rise in the torque density is most significant.  

The total integrated net torques at 1 and 2 AU in the model disk 
are greater than the total torques for an equivalent isothermal disk.  
This is because the temperature decreases with $z$ up to 
the photosphere in the 
model disk, so the density falls off more than in 
a vertically isothermal disk, as shown in \figref{diskprof}.  
This leads to greater sampling 
of the torque at the disk midplane relative to the photosphere.
As the distance from the star increases, the 
vertical temperature profile flattens, and the density 
profile rises relative to the profile of the isothermal disk.  
Contributions to the total torque from the upper layers rise 
relative to contributions from the midplane, effectively softening 
the total torque as a whole, and the total net torque of the model 
disk falls below that of the isothermal disk.  

\begin{figure}
\plotone{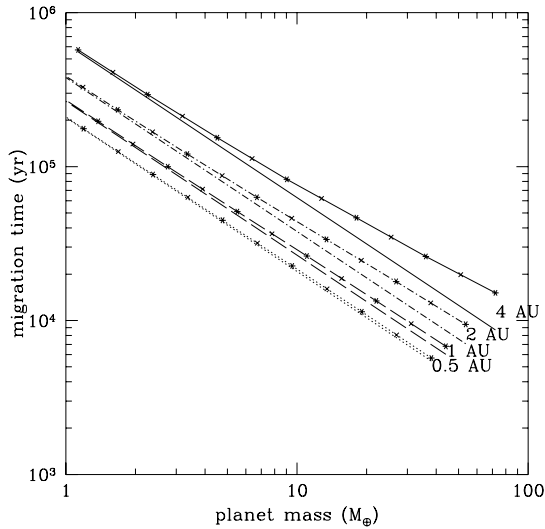}
\caption{\label{timescales}Comparison between migration timescales 
in the disk-planet system in the presence of temperature 
perturbations and the absence of temperature perturbations.  
We plot the migration timescale versus planet mass at 
0.5 (dotted line), 1 (dashed line), 2 (dot-dashed line) and 4 (solid line) 
AU, showing the perturbed and unperturbed times as lines 
with and without points.  (The six-pointed stars represent 
calculated models, the crosses are interpolations for illustration.)
}
\end{figure}

At 1 AU and inwards, the effect of planet-induced temperature 
variations are small.  
However, as the distance increases to 2 and 4 AU, 
planet-induced temperature variations become more effective at 
changing the torque from Lindblad resonances.  These results are 
shown in \figref{timescales}, where we plot migration time scales 
versus planet mass at 0.5, 1, 2, and 4 AU as dotted, dashed, dot-dashed, 
and solid lines, respectively.  
The lines with and without points show migration time scales in the
model disk with and without considering local temperature
perturbations, respectively.  At 0.5 AU, the migration timescales are
changed very little by the temperature perturbations.  Further from
the star, the departures from the unperturbed disk become more and
more pronounced.  This is, in part, due to the vertical density
profile, which weights the torques at the midplane more heavily at
closer distances to the star, as discussed above.  But as can be 
seen from Figures \ref{t1}, \ref{t2}, and \ref{t4}, the torques at 
the midplane are also more affected with increasing distance.  This 
is because the total optical depth of the disk decreases with 
distance as the surface density falls off.  This increases 
shadowing and illumination effects at the midplane, which has a 
greater effect on the net Lindblad torques there.  

\begin{figure}
\plotone{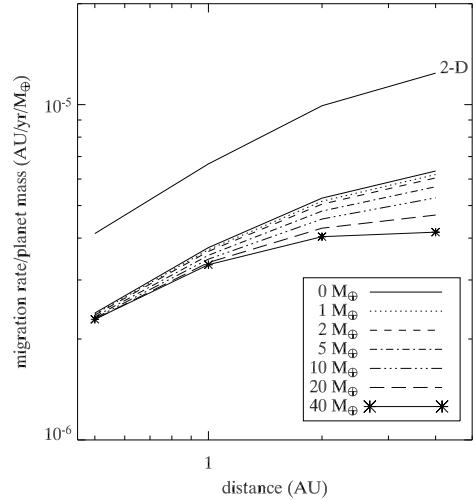}
\caption{\label{mig_rates}Migration rates for the model disk-planet system 
scaled by the planet mass.  The upper solid line labelled ``2D'' is 
the migration rates for an equivalent two-dimensional disk 
using the formulation of \citet{wardA}.  The lower lines are 
the migration rates calculated for the full three-dimensional 
disk including temperature perturbations from a planet with 
mass indicated in the legend.  The solid line, marked 0 M$_{\oplus}$
in the legend, shows migration rates for the disk in the absence of 
any planet-induced temperature perturbations.  
}
\end{figure}
We re-plot these results in terms of migration rate per Earth mass 
versus distance from the star in \figref{mig_rates}.  
Given the grid of planet distances and masses we have sampled, 
we interpolate to find the migration rate for selected planet masses 
and plot the variation of migration rate with distance for 
different planet masses, as indicated by the different lines. 
In the absence of perturbations to the disk, 
the migration rate scales with the mass of the 
planet, so this is good way to concisely summarize the effects of 
local temperature variations.  Close to the star, even planets 
at the gap-opening threshold reduce the migration rate 
by only a few percent.  However, as the distance increases, 
planet-induced temperature variations become more and more effective 
at lowering the migration rate, up to $43\%$ at 4 AU.  
For comparison, we also plot the migrations rates for a 
comparable two-dimensional disk, which gives migration rates 
of about twice as fast as the unperturbed three-dimensional disk.  
For this model disk, the overall trend is that planetary migration 
slows as the planet moves closer to the star rather than staying 
constant, although this deceleration is not sufficient to keep 
planets from migrating right into the star.  

\newpage
\section{Discussion}
\label{discussion}

Although the density of the disk is highest at the midplane, 
the contributions to the Lindblad torques at the midplane 
do not solely determine the total net migration torque.  
The finite thickness of the disk is also important because 
it both modifies the perturbing potential and shifts the 
resonance positions.

\begin{figure}[htbp]
\plotone{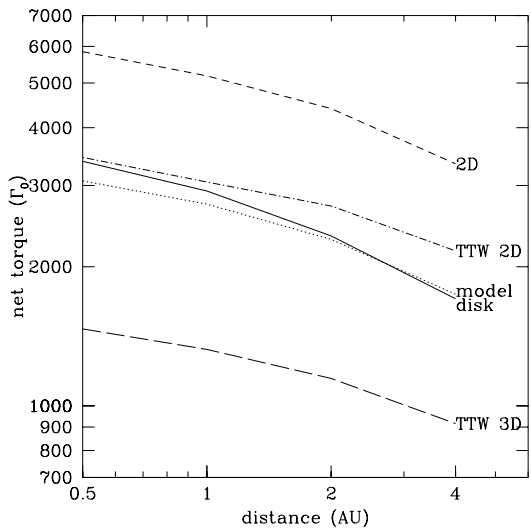}
\caption{\label{comprates}Comparison of migration torques between 
different disk models vs.~radius.  All disk models have 
the same midplane density and surface density as the model disk.  
We show the three-dimensional model disk ({\em solid line}), 
a vertically isothermal model ({\em dotted line}), and a 
two-dimensional disk ({\em short dashed line}).  We also show for 
comparison the calculation done by \TTW{} for a 
three-dimensional ({\em long dashed line}) and 
two-dimensional ({\em dot-dashed line}) disk. 
}
\end{figure}

For our model disk, 
we find that including the thickness of the disk reduces the 
migration rate by about a factor of two from a two-dimensional disk.  
By comparing the model disk with an equivalent isothermal disk, 
we find that including the vertical softening of the perturbing 
potential is, in fact, the primary effect, and that the 
vertical temperature structure has a modest effect.  
We show how the torques in these model disks compare to each 
other in \figref{comprates}.  The model disk is shown as 
a solid line, the two-dimensional disk is the short-dashed line, 
and the isothermal disk as a dotted line.  

\TTW{} also find that the net Lindblad torques in 
their calculation of a three-dimensional isothermal disk 
are a factor of $\sim2-3$ less that of the equivalent two-dimensional 
disk.  These torque calculations are also plotted in \figref{comprates}
as a long dashed line for the three-dimensional disk and a 
dot-dashed line for the two-dimensional disk. 
While there is a mismatch between the \TTW{} models and our model 
disk, the results are still qualitatively the same in showing 
how including the vertical structure of the disk affects 
the total torque on the planet.  

Planet-induced temperature variations can reduce the migration rate 
relative to that of the unperturbed disk by up to a factor of about 2 for 
planets at the gap-opening threshold at around 4 AU.  
Migration rates are less affected closer to the star 
because of increased viscous heating at the midplane.  
In general, the midplane of the 
disk contributes most to the resonant torques since  
temperature variations are minimal both because of the 
high optical depth to the stellar irradiation and viscous heating 
at the midplane.  Viscous heating and stellar irradiation 
behave as $r^{-3/4}$ and $r^{-1/2}$, respectively, so viscous 
heating increases faster with decreasing radius than stellar irradiation.  
Midplane temperature perturbations are therefore washed out 
by viscous heating closer to the star.  
This is why the presence of a planet has little effect at 
0.5 AU, but is highly effective at 4 AU.

The results of \citet{menougoodman} also show that the thickness 
of a three-dimensional disk accounts for a significant
increase in migration timescales over those of two-dimensional disks.  
They consider an $\alpha$-disk model similar to ours, 
and find that opacity changes in the disk create local 
regions of increased migration timescale, or decreased migration rate, 
because of local variations in the temperature gradient.  
The spikes in the migration timescale that result from 
local changes in the temperature gradient close 
to the star ($\lesssim1$ AU) may be softened by shadowing 
effects, since these local effects may dominate over the 
large-scale temperature gradient.

As a planet migrates inward, the gap-opening threshold mass decreases, 
so the planet may begin to open a gap and undergo Type II migration 
rather than Type I migration.  This is a feature of 
any disk model with a flared structure, where $h$ increases 
faster than $r$.  The gap-opening threshold is where 
$\rhill \equiv (m_p/3M_{\star})^{1/3}r = h$.  Rewriting this 
equation, the gap-opening threshold mass is 
\begin{equation}
m_p = 3 M_{\star}\left(\frac{h}{r}\right)^3.  
\end{equation}
If $d\log h/d\log r>1$, then the gap-opening mass increases with $r$.  
In Type II migration, the planet remains fixed with respect to the 
disk instead of moving through it, as in Type I migration.  
As the disk is accreted onto the star, the planet moves with it 
in the viscous time scale.  
This is an altogether different mechanism for slowing migration. 
However it still suffers 
from the problem that as long as the disk exists, the planet 
will migrate together with the disk and could still fall into 
the star.

\section{Summary and Conclusions} 
\label{summary}

We have analyzed the effect of local disk temperature variations 
on Type I migration rates of planets embedded in protoplanetary 
disks.  We include both the vertical structure of the disk 
as well as temperature perturbations caused by the presence of the 
planet itself.  

In comparison to the results of \citet{wardA} for a two-dimensional 
disk model, the primary difference in migration rates is 
caused by including the vertical thickness of the disk.  
This is because the perturbing potential is diluted by the 
vertical distance from the midplane.  When we integrate the 
total torque at Lindblad resonances over the vertical extent of 
the disk, we find that this 
reduces the migration rate by about a factor of two.  
This is consistent with calculations by \TTW{}, who 
find that the vertical dependence of the perturbing potential 
reduces migration rates by a factor of $\sim2-3$ in 
their model disk.

We estimate the effect of including the 
vertical temperature and density profile of the model disk 
by comparing it to an equivalent vertically isothermal 
disk and find that the effect on the migration rates is modest.  
This is in part because the torque is weighted by disk density, 
which is highest at the midplane, and also because of 
the rapid decrease in torque density with disk height.  
Therefore disks with the same midplane temperature and 
density profiles will yield similar migration rates.  

When we include the effects of radiative transfer on the 
perturbation caused by the planet on the disk, we find 
that the resulting temperature perturbations can further 
decrease the migration rate by up to about a factor of two, 
depending on the planet's mass and position.  
As the planet increases in mass, so does the temperature 
variation, which results in a general decrease in the migration rate. 
The effect is more prominent in planets further from the 
star, because viscous heating decreases faster than 
stellar illumination with increasing distance.  Viscous 
heating evens out temperature variations, reducing their 
impact on Lindblad torques.  Thus, the planet at the gap-opening 
threshold at the farthest distance in our sample is the 
most effective at reducing its own migration rate.  

Our calculation of migration torques given the temperature structure
in the photosphere of the disk shows a limiting case, since we assume
that only the local temperature of the disk is relevant, and that the
azimuthal dependence on temperature is unimportant.  This is 
justified by the fact that the perturbing potential is 
strongest closest to the planet, but in principle the azimuthal 
variation of the temperature could dilute the effect of 
local temperature variations on migration rates.  

Another limitation of our planet-disk model is that we have 
not considered the non-linear perturbation to the 
density structure within a few Hill radii of the protoplanet.  
Since the tidal torques from Lindblad resonances 
scale with disk density, this effect may be very important.  
Numerical simulations universally show that spiral structure 
develops quickly around planets embedded in disks.  
Radiative transfer from stellar illumination upon these structures 
can affect the planet and disk both directly through heating, 
and indirectly by changing the positions and strengths of 
the Lindblad resonant torques.  

Still, our model puts interesting constraints on the effect of 
the temperature and density structure on Type I migration.  
We have shown that shadowing and illumination at the surface of a 
protoplanetary disk can slow migration.  If the perturbations 
are large enough, they might be able to halt or even reverse 
Type I migration, and save planet embryos from the 
``Shiva catastrophe'' \citep{wardB}.  Perhaps this can explain 
why the frequency of giant planets around solar-type stars is 
relatively high, despite the short timescales predicted 
for Type I migration.  

\acknowledgements

We thank Kristen Menou and our anonymous referee for their 
helpful comments in the preparation of this paper.

\bibliographystyle{apj}
\bibliography{../planets}

\end{document}